\documentclass[graybox]{svmult}

\usepackage{hyperref}
\usepackage{color}

\usepackage{mathptmx}       
\usepackage{helvet}         
\usepackage{courier}        
\usepackage{type1cm}        
\usepackage{graphicx}        
\usepackage[bottom]{footmisc}
\usepackage{chicago}
\usepackage{algorithm}
\usepackage{algorithmic}

\begin{document}
\title*{Communities as Well Separated Subgraphs \\With Cohesive Cores: Identification of\\ Core-Periphery Structures in Link Communities}

\titlerunning{Communities as Well Separated Subgraphs With Cohesive Cores} 
\author{Frank Havemann, Jochen Gl\"{a}ser, and Michael Heinz}
\institute{Frank Havemann and Michael Heinz \at Institut f\"{u}r Bibliotheks- und Informationswissenschaft, Humboldt-Universit\"{a}t zu Berlin, Germany, \email{frank.havemann/michael.heinz@ibi.hu-berlin.de} \\ Jochen Gl\"{a}ser \at Center for Technology and Society, TU Berlin (Germany)\at  {\small --------------------------------------------------------------------------------------------------------------\\Submitted version accepted for the 7th International Conference on Complex Networks and Their Applications, December 11--13, 2018, Cambridge, UK}. \\Revised version at \url{http://141.20.126.227/~qm/papers/} }
%
%
\maketitle

\abstract{
Communities in networks are commonly considered as highly cohesive subgraphs which are well separated from the rest of the network. However, cohesion and separation often cannot be maximized at the same time,  
which is why  a  compromise is sought by some methods. When a compromise is not suitable for the problem to be solved
it might be advantageous to separate the two criteria. In this paper, we explore such an approach by defining communities as well separated subgraphs which can have one or more cohesive cores surrounded by peripheries. We apply this idea to link communities and present an algorithm for constructing hierarchical core-periphery structures in link communities and first test results.
}

\section{Introduction}
\label{sec:1}
Communities in networks are commonly considered as subgraphs with dense internal but sparse external connections (cf., e.g., \citeN{Girvan2002community}, \citeN{bagrow_local_2005}, and \citeN{fortunato2010community}). 
In other words, a community should be a highly cohesive subgraph which is well separated from the rest of the network. Maximum cohesion of nodes is reached in fully connected subgraphs (cliques), maximum separation for subgraphs without external connections (components).  In many practical cases the two essential features of communities cannot be maximised both at the same time, which is why a compromise 
is sought by some methods
for the construction of communities (for reviews of community finding, see \citeN{fortunato2010community}, \shortciteN{Xie:2013:OCD}, and \citeN{amelio_overlapping_2014}). 
When a compromise is not appropriate for the problem to be solved,
it might be advantageous to separate the two criteria. In this paper, we explore such an approach by defining
 communities  as well separated subgraphs which can have one ore more cohesive cores surrounded by less cohesive peripheries. Due to the size bias of cohesion measures we favour separation as the defining feature.  Methods for finding core-periphery structures were reviewed by \shortciteN{csermely_structure_2013}. 

We apply this idea to link communities in networks. Link clustering was introduced by \citeN{evans2009line} and by \shortciteN{ahn2009link}. The aim of our paper is to operationalise the argument for separating the evaluation of cohesion and separation for link communities, and to propose an algorithm that identifies core-periphery structures in well separated link communities. In Section \ref{sec:2} we discuss the conceptual  problems of simultaneously maximising cohesion and separation. 
In Section \ref{sec:4} a method for determining core-periphery structures of link communities is derived. In Section~\ref{sec:5} it is applied to results of a local link clustering exercise \shortcite{havemann_memetic_2017} and  to the karate-club network \cite{zachary1977information}.

\section{Cohesion and Separation}
\label{sec:2}
If communities in networks are considered as highly cohesive and well separated subgraphs, all cliques without external links are ideal but trivial and very rare communities of nodes. In nearly all cases we have to content ourselves with imperfect communities. In this section, we discuss three problems of community construction, namely (A)  the existence of real-world problems for which the maximisation of cohesion is likely to create artifacts, (B) the necessity to compromise between maximising cohesion and separation for all other real-world problems, and (C) the size bias of most measures of cohesion. 

 \textbf{(A) Maximisation can produce artifacts:}
Some real-world problems are represented by communities that contain the boundaries of other communities. This is the case when communities form a hierarchy, i.e.\ when larger communities contain smaller ones. In this case, the smallest communities of a hierarchy can be rather cohesive but  supercommunities only if their subcommunities are not very well separated \cite{ravasz_hierarchical_2003,rezvani_fach:_2018}.
A second case is communities overlapping pervasively (i.e.\ not only in boundary nodes). Here, too, boundaries of one community run through another, thereby lowering the cohesion and violating the demand that communities should be hard to split \shortcite{kannan_clusterings:_2004,leskovec_empirical_2010,yang_defining_2013}. Applying a cohesion-maximising algorithm to these problems might lead to important communities being excluded from consideration, or to artefactual communities being included.

\textbf{(B) Compromising:} The best way to compromise between cohesion and separation may be difficult to determine. This problem occurs especially with approaches that evaluate single communities. If whole networks are partitioned into disjoint communities the compromise is often built into the algorithm and cannot be considered separately (e.g. in the case of modularity-maximising algorithms, \citeN{Newman2004finding}). If an algorithm evaluates single communities, cohesion and separation are unlikely to be maximal for the same subgraph, which necessitates a compromise. This raises the question how such a compromise should look like. \citeN{pizzuti2009multi} introduced a bi-objective optimisiation  which allows to choose an appropriate compromise between the two features. She used a genetic algorithm to maximise internal and to minimise external connectivity of a partition's communities. 
\shortciteN{kannan_clusterings:_2004} proposed to solve this problem by setting a minimum level for one feature and  maximising the other under this minimum condition.  However, it depends on the real-world problem to be if and how a suitable compromise can be found.

\textbf{(C) Size bias of cohesion measures:} One of the major -- and so far underappreciated -- problems of community construction is the size bias of most cohesion measures. In general, \textit{global}  cohesion of a  set can be measured by the ratio of the number of directly connected element pairs to the theoretical maximum of this number.\footnote{For the cohesion of nodes in monopartite topological graphs this ratio equals \textit{link density} which is maximal for cliques. For link communities, we derived an analogue to link density named \textit{connectedness density} of links (see App., p. \pageref{app.2}). It is maximal for star subgraphs.}
In imperfect communities, links may be so unevenly distributed that the communities contain well separated and cohesive subgraphs. This is why some authors demand that in addition to be highly cohesive, a community should be ``hard to split'' and measure cohesion by internal conductance, i.e.\ the minimal conductance of all possible splits \shortcite{kannan_clusterings:_2004,leskovec_empirical_2010,yang_defining_2013}. Although internal conductance is expected to have no size bias, its calculation depends on the identification of the best split, which creates significant problems for community construction. Furthermore, this demand cannot be upheld for the problems described under (A) above.
For some practical problems it is sufficient that members of communities have a high \textit{local} cohesion; cf., e.g., \shortciteN{xu_scan:_2007}. Local cohesion of nodes can be measured, e.g., by their degree or the local clustering coefficient.
Most cohesion measures based on network topology solely are scaling with size in sparse networks.
Link density tends to be smaller for larger subgraphs (cf.\, \citeN{schaeffer_graph_2007}, p.\,50).\footnote{Similar to the link density of nodes, connectedness density of links scales with size: larger link communities tend to have lower values.}
When average internal degree is used to evaluate the cohesion of a community the opposite size bias is observed. 
For scale-free networks the average clustering coefficient decreases with size \cite{ravasz_hierarchical_2003}.

A further option to measure cohesion seems to be to relate the sum of internal degrees $k_\mathrm{in}(C)$ of nodes in $C$ to the sum of their total degrees $k(C) = k_\mathrm{in}(C) + k_\mathrm{out}(C)$. This ratio equals the probability that a random walker found in node community $C$ does stay within $C$ in the next step and is therefore called  \textit{persistence probability} \cite{Piccardi2011findingPLOS,rossa_profiling_2013}:
 \begin{equation}
P_\mathrm{pers}(C) =  \frac{\sum_{i \in C} k_i^\mathrm{in}(C)}{\sum_{i \in C} k_i(C)} = \frac{k_\mathrm{in}(C)}{k(C)}.
\end{equation}
$P_\mathrm{pers}(C)$ appears to measure cohesion of nodes but is insensitive to  the distribution of connections. Two subgraphs with the same number of external and of internal links have the same persistence probability 
but can have a rather different cohesion of nodes measured by their link density or their internal conductance. 
The random walker needs many internal and a few external links to walk within $C$ for a while but the internal structure is not relevant. For example, $C$ can also be a chain of nodes or even be disconnected. This means, that persistence probability $P_\mathrm{pers}(C)$ measures separation  rather than cohesion as defined here. With increasing external degree  $k_\mathrm{out}(C)$ persistence probablity decreases. This can be made explicit when we rewrite it:
$P_\mathrm{pers}(C) = 1 - k_\mathrm{out}(C)/k(C)$, 
where $k_\mathrm{out}(C)/k(C)$ equals the probability of a random walker found in $C$ to leave the community in the next step, also called \textit{escape probability} and denoted here by $P_{\mathrm{esc}}(C)$ \shortcite{fortunato2010community}, cf.\,App., p.\,\pageref{App}.  

\citeN{Piccardi2011findingPLOS} pointed to the fact that $P_\mathrm{pers}(C)$ is related to the definition of \textit{communities in the weak sense} given by \shortciteN{Radicchi2004defining} with the criterion $k_\mathrm{in}(C) > k_\mathrm{out}(C)$. If $P_\mathrm{pers}(C) > 1/2$ then this criterion is fulfilled. The strong definition of communities demands that each node has to have more internal than external links. If $k_\mathrm{out}(C)$ is small, both definitions tolerate communities which can be split. Consider, for example, a subgraph comprising  two 4-cliques with one external link per clique and one link between both cliques. Both cliques are also communities in the strong and weak sense.
Escape probability $P_{\mathrm{esc}}(C)=k_\mathrm{out}(C)/k(C)$ is used in cut based measures of separation as  \textit{conductance} and \textit{normalised cut}. In Appendix, p.\,\pageref{App}, we discuss these measures and also \textit{normalised node-cut}, a measure of separation for link communities proposed by us recently \shortcite{havemann_memetic_2017}. 

Our discussion here is limited to topological networks. We transcend network topology if a suitable measure of node distance can be defined. Then global cohesion can be defined as some aggregate of distances between a subgraph's nodes. As a suitable measure we consider a distance which is not maximal for unlinked nodes. If all unlinked nodes have the same distance the ends of a long chain would have the same distance as two nodes in the chain with a third node between them. Distance should also not strongly depend on the  position of individual links which is the case for length of the shortest path and its derivatives. 

In summary, maximising cohesion of communities in topological graphs is difficult when (A) maximising cohesion, (B) compromising between cohesion and separation, or (C) a size bias of cohesion measures may lead to the disregard of subgraphs that are important to solving specific real-world problems. In these cases, separation and cohesion can be measured for different objects.
This can be achieved if we introduce the notion of cohesive community cores. 
Like whole networks, communities can have a core-periphery structure \cite{yang_overlapping_2014,kojaku2017findingPRE}. A cohesive core can be linked to many peripheral nodes, which means that it is not well separated. Separation can be improved by including the  core's periphery into the community which simultaneously diminishes its internal cohesion. 
In order to realise separate measurements, we propose to consider communities in networks as well separated connected subgraphs and to reserve the feature of high cohesion for community cores. In other words, we propose to change the common notion of communities in networks in those cases where it is unnecessary but we rescue separation and cohesion as aims of optimisation. In the language of \shortciteN{kannan_clusterings:_2004} we maximise the subgraph's separation while the minimum condition for its cohesion is its connectedness. We then look for cohesive cores of communities. We propose to define a cohesive core and its periphery  not in absolute terms but as a sequence of nested subgraphs with decreasing cohesion.

\section{Core-Periphery Structures in Link Sets}
\label{sec:4}

There are several methods for finding cohesive cores of graphs or node communities \shortcite{borgatti2000models,zhang_identification_2015}. We construct core-periphery structures in link communities as nested subgraphs with decreasing connectedness density  (see App., p. \pageref{app.2}). We start from subgraphs with local maxima of local density of links which are sufficiently distant from other subgraphs with local maxima analogously to methods for node-community finding proposed by \shortciteN{liu_novel_2017} and by \shortciteN{wang_locating_2017}. The simplest way to translate local node density used by these authors into the world of link clustering is to define local  density of links as the number of neighbouring links attached to a node. 
Thus, local  density of links equals local node density. Large stars as link sets with maximal connectedness density have then also a high local  density of links. Therefore, we start from the largest star of a link community $L$ for constructing its core-periphery structures. We apply the same definition of local density but differ from \shortciteN{wang_locating_2017} and \shortciteN{liu_novel_2017}, who construct disjoint clusters of nodes, by  clustering stars as link sets and allowing for overlap in nodes and links.\footnote{Both groups use community centres as seeds for a local expansion. \shortciteN{wang_locating_2017} apply a greedy algorithm maximising persistence probability $P_\mathrm{pers}(C)$ which they assume to measure subgraph density (cf.\,discussion above).\label{Wang} \shortciteN{liu_novel_2017} propagate the labels of centres to nodes. For bipartite networks our algorithm can also be seen as an adaption of the method proposed by \shortciteN{carmi_partition_2008} for partitioning a network into basins of attraction to hubs. \citeN{da2004hub} also used high-degree nodes (hubs) as centres of communities which he constructed by a simple expansion process starting from a predefined number of hubs.}
Different from us, \shortciteN{zhou_density_2017} directly translate local density from the world of node clustering to links to obtain a link-clustering method. 

\begin{algorithm}[t]
\caption{Pseudocode  of CPLC (cores and peripheries of link communities)}
\label{pseudcode-CPLC}
 \begin{algorithmic}
 \STATE     
\STATE all stars of a (sub)graph are centre candidates; 
\STATE the largest star is a centre (select one randomly from two or more largest stars);
\STATE remove it from set of centre candidates;
\STATE initialise its town with the centre itself;
\WHILE {there are centre candidates}
	\IF {the largest centre candidate $j$ shares at least one link \OR more than $qk^\mathrm{in}_j$ of its outer nodes with any town}
	\IF {there is only one such town}
   \STATE centre candidate $j$ is united with the town; 
   \ELSE
\STATE the links of $j$ to a town are united with the town;
\STATE the remaining links of $j$ are united with all towns with overlap;
   \ENDIF 
    \ELSE
   \STATE  candidate $j$ becomes a new centre;
   \STATE initialise its town with the centre itself;
 \ENDIF
 \STATE remove stars which have all links within towns from set of centre candidates;
 \STATE skip the next centre candidate(s) which share all links to towns with towns;
\ENDWHILE
 \end{algorithmic}
\end{algorithm}

Our aim is to determine hierarchical core-periphery structures (named \textit{towns}, for short) in a given connected subgraph induced by link set $L$.  We  define a town as a hierarchy of stars where two stars are never indirectly connected with each other via smaller stars only. Two stars are directly connected if they share a link or one of their outer nodes. A star is connected to a town if it shares a link or a minimum number of outer nodes with the set of town stars of equal or larger size; otherwise it becomes the centre of an independent town. The minimum number of outer nodes is determined by a resolution parameter $q$ with $0 \le q < 1$ which is used as a minimum threshold of relative overlap for a star to be attached to a town. If $q = 0$ one common node of star and town is enough to unite both link sets. If $q = 1/2$ more than half of the star’s outer nodes have to be already inside the town. If there are two or more towns a star is connected to it is split and its parts are attached to these towns. 

The algorithm for finding cores and peripheries in link communities  (CPLC algorithm) can be described as follows (cf. Algorithm \ref{pseudcode-CPLC}).
All star subgraphs of the community are ranked with regard to their size. To construct a town $T$ it is initialised by the largest star. The next star on the rank list  is attached to  town $T$ with node set $N_T$ if the number of the star's outer nodes shared with the town fulfils the minimum condition given by resolution parameter $q$: 
$|\mathrm{adj}(j) \cap N_T| > qk^\mathrm{in}_j$, 
where $\mathrm{adj}(j)$ denotes the adjacent nodes of the star's central node\,$j$. A direct link between two star centres is also a sufficient condition for being included in the town. If a star could be united with two or more existing towns then we  add to each town its links with this town. Its remaining links are united with all towns involved.
Then we delete all (mostly small) stars from the list of candidates which now have no links outside any town. 
We  skip candidate stars that share all links to towns with these towns. We found this feature useful in our experiments with link communities in a nearly bipartite network where different kinds of nodes can be centres of candidate stars. Skipping these stars does also work in the unipartite karate-club network.   

The number of towns obtained depends on resolution $q$. Instead of voluntarily setting parameter $q$ we  explore its whole range.   
We start with minimal resolution $q = 0$ and obtain a structure of the subgraph with the minimal number of towns. As long  as $q < 1$ we then recursively increase $q$ to a value at which it is possible to obtain at least one town more. Therefore, the next threshold is taken from the smallest portion of nodes shared by a star and a town with which the star was united. This guarantees that in the next run of CPLC this star and all stars with the same relative overlap to any of the towns are not united with these towns.


To select  
resolution levels at which  
relatively well separated towns are obtained
we calculated normalised node-cut $\Psi(L)$ of link set $L$ for each town (cf. Equation \ref{eq:Psi} in App.; here $E$ is the link community analysed). Because towns are not optimised with regard to separation we also calculated function $\Psi$ for $L$ subtracted by all links in overlaps between towns. For each town we choose the better (lower) of both values and evaluated the resolution level with the worst (largest) $\Psi$ of any town. If there are two or more levels with same worst $\Psi$ and same number of towns, we selected the lowest level with minimum link overlap between towns.

\section{Experiments}
\label{sec:5}
\begin{figure}[b]
\sidecaption
\includegraphics[scale=.56]{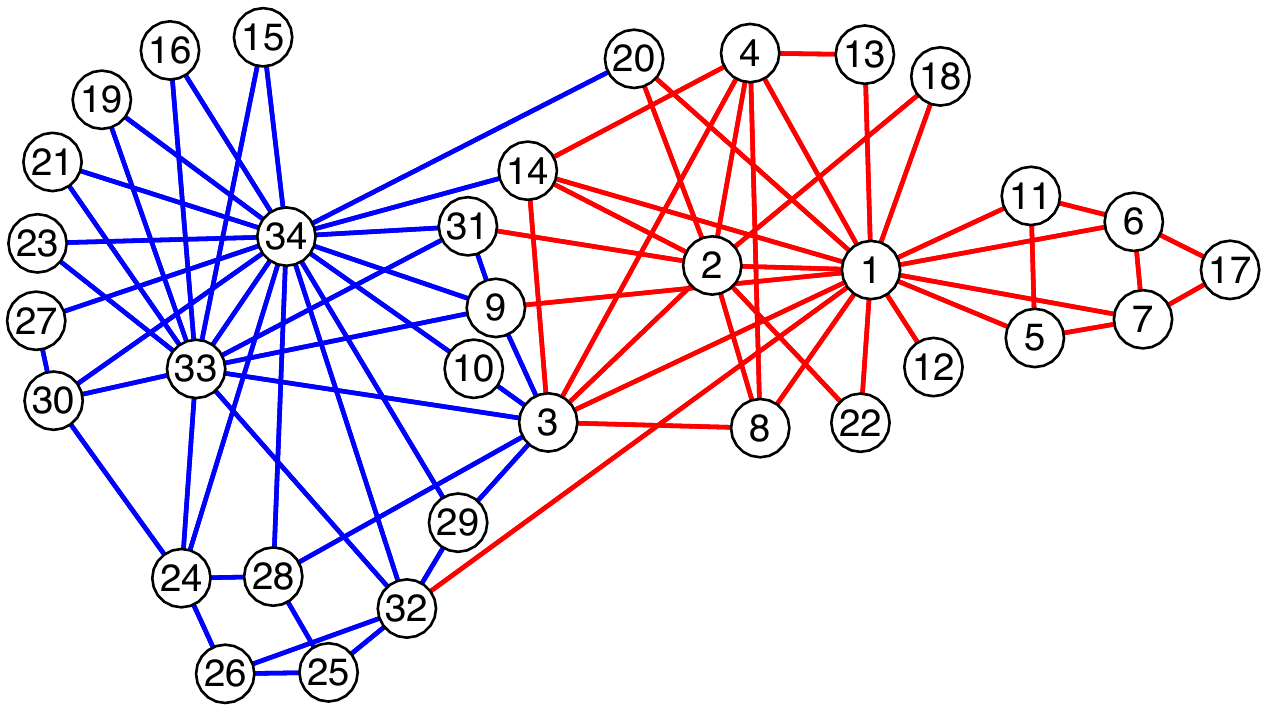}
%
%
\caption{Network of 34 karate fighters organised in a club and observed by Zachary (1977) in their connections outside the club. The split into red and blue links has minimal $\Psi$-value (cf. text).}
\label{fig:1}       
\end{figure}
The karate club analysed by \citeN{zachary1977information} has only one town for lowest resolution level $q=0$ with node 34 as the centre. We obtain two towns for resolution $q \ge 1/4$ with centres 1 and 34. For $q \ge 4/9$ their link overlap reduces to three links: $(3, 9)$, $(3, 14)$, and $(9, 31)$. Besides the nodes of these three links, the two towns overlap in nodes 20 and 32. Thus, the two towns are compatible with the final splitting of the karate club due to conflicts between the two leaders 1 and 34. For the town with centre node 1, $\Psi= 0.1770$  and $\Psi=0.1723$ for the other town. A better value $\Psi=0.1638$ (in fact the best we have found) can be obtained for a disjoint link splitting of the karate-club network where the towns' link overlap is split up  (s.\,Fig.\,\ref{fig:1}).

\begin{figure}[t]
\sidecaption
\includegraphics[scale=.59]{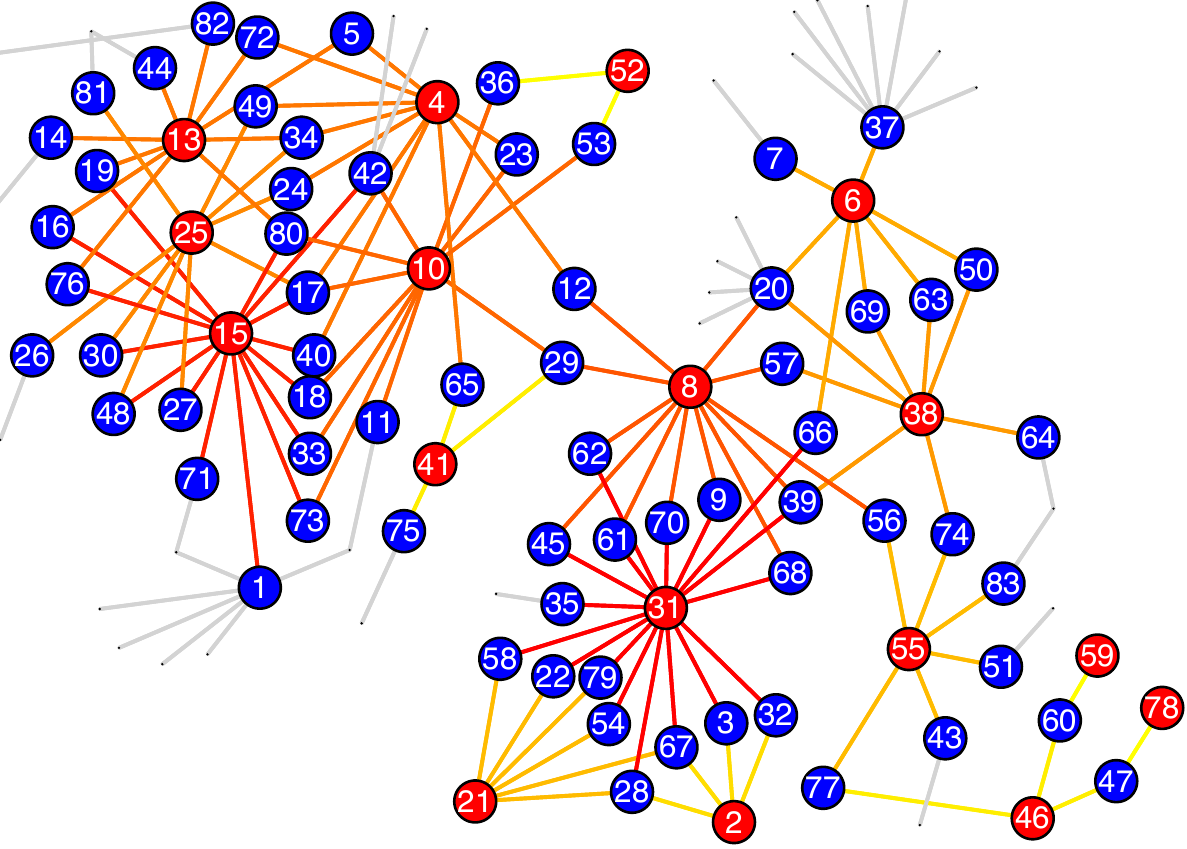}
%
%
\caption{Community of citation links between 17 papers (red nodes) and 66 cited sources (blue nodes). Links are coloured from red to yellow according to the size of the largest star they belong to. Grey links are not community members but neighbours. A minimal $\Psi$-value within the subgraph is obtained if the two blue nodes 12 and 29 in the centre of the subgraph are cut through  (cf.\,text).}
\label{fig:2}       
\end{figure}

Searching for link communities with locally minimal $\Psi$-values in a nearly bipartite network of 14,770 papers published 2010 in astronomy and astrophysics journals (including also geophysical papers) and their cited sources 
we found 127 overlapping link communities which cover the network and form a poly-hierarchy \shortcite{havemann_memetic_2017}. All sources cited only once were omitted. Fig.\,\ref{fig:2} shows the small community  \tens{h80} which is an example of a well separated subgraph ($\Psi = 0.0458$) but it is not very cohesive (it can be split into two subcommunities).  Here we have two towns already at zero resolution. Increasing it to $q=1/10$ causes their overlap to decrease from 16 to 2 links ($(29, 41)$ and $(41, 75)$) and $\Psi$ of towns within the subgraph reaches a minimal value of 0.0187 if the two overlapping links are deleted from the town on the right-hand side. For $q \ge 1/3$ we find solutions with four and more towns but relatively bad separation (worst $\Psi \ge 0.1313$). The centres of the two towns are the (deep red) large stars with central nodes 15 and 31. The two towns correspond to two subcommunities we had found with our search algorithm. The ten papers in the town on the right-hand side of the subgraph deal with lightnings and similar electromagnetic phenomena in the atmosphere, the seven papers on the left-hand side mainly deal with  effects of seismic activities measured in the ionosphere.

The number of stars and the number of towns both increase with the number of links $m$. We therefore expect run-time of CPLC to scale with  $m^2$ which is confirmed by experiments with 151 communities found in the astrophysical citation network (including the 127 communities mentioned above and the whole network with $m=536,020$). Due to space limitation we cannot present further statistics of  results. 

\section{Summary and Discussion}
\label{sec:6}

If a real-world problem is likely to be represented by a hierarchy of communities or overlapping communities or by communities of varying and unknown sizes, or if it is difficult to determine the best compromise between cohesion and separation, it seems advantageous to separate the maximisation of cohesion and separation. In this paper, we propose a strategy that starts from communities as well separated subgraphs and identifies cohesive cores of such subgraphs. We applied this strategy to the analysis of link communities. 
To determine core-periphery structures as hierarchically nested subgraphs with decreasing cohesion in a link community we start from local maxima of local link density, i.e., from the largest stars. The examples presented here demonstrate that the algorithm we have tested is able to separate core-periphery structures (\textit{towns}) if there are two or more such structures in a (sub)graph. 
 Our next task is to evaluate each town  with regard to their distinctness. If all stars have nearly the same size it would be difficult to speak of a hierarchy with centre and periphery.
A further task is to assess the correspondence of core-periphery structures with the real-world problem of research topic detection. Towns of communities are expected to correspond to sub-topics of the larger topic represented by the community.


Beside high cohesion, another often assumed feature of cores is their network centrality \shortcite[p.\,94]{csermely_structure_2013}, e.g., indicated by low average distance to peripheral nodes. We are also interested in non-central cores of link communities. 

Towns of the whole graph can be used as seeds for local link clustering. Towns of communities found can recursively serve as seeds for finding smaller communities.

\section*{Appendix}

\section*{Cohesion of Link Sets}
\label{app.2}
Let $k_i^\mathrm{in}(L)$ be the number of links in set $L$ attached to node $i$, also called its internal degree. The number of links  in $L$  a link $(i, j)$ is connected to equals $k_i^\mathrm{in}(L) + k_j^\mathrm{in}(L) - 2$. For the total number of (ordered) pairs of directly connected links in $L$ we find
\begin{equation}
N(L) = \sum_{(i, j) \in L} (k_i^\mathrm{in}(L) + k_j^\mathrm{in}(L) - 2) = \sum_{i=1}^n k_i^\mathrm{in}(L)\cdot(k_i^\mathrm{in}(L) - 1). 
\end{equation}  
In the sum each node $i$ occurs $k_i^\mathrm{in}(L)$ times with connections from one link to $k_i^\mathrm{in}(L) - 1$ others. $N(L)$ is an absolute measure for cohesion of link sets. It is not maximal if the $|L|$ links form a clique of nodes. Indeed, for a clique of four nodes connected by six 
links we have $N(L) =  24$. If the six links form a star we obtain a higher value $N(L) =  30$. In the star graph all links are directly connected but in cliques of at least four nodes not. This corresponds to the fact that the line graph of a star is a clique. 
If $L$ has the form of a star and $c$ denotes its central node then $k_c^\mathrm{in}(L) = |L|$. For all other nodes we have $k_i^\mathrm{in}(L) = 1$, i.e., $N(L) = |L|(|L| - 1)$. 
Link sets are maximally connected if all their links are directly connected by a node, i.e.,  stars  are maximally connected link sets. 

We can define a relative measure of cohesion of link sets by relating absolute node connectedness of links $N(L)$ to its maximum reached by stars. That means, as \textit{connectedness density} $D$ of a link set $L$ we can define  
\begin{equation}
D(L) = \frac{\sum_i k_i^\mathrm{in}(L) (k_i^\mathrm{in}(L) - 1) }{ |L|(|L| - 1)}. 
\end{equation}
This measure is useful for both one-mode and also for two-mode networks (where are no cliques). In both types of networks stars are the most cohesive link sets.  Analogously to this measure, link similarity as defined by \shortciteN{ahn2009link} is not maximal for all link pairs in a clique of $n$ nodes but in a star of $n(n - 1)/2$ links.  

\section*{The Random Walker and Separation of Communities}

\label{App}
Supposing that a random walker is on any node in set $C$ (in an unweighted and undirected network), his probability to be on node $i$ equals $k_i/ \sum_{i \in C} k_i.$ He leaves $C$ in the next step with probability $k_i^\mathrm{out}(C)/k_i$. Then his probability to leave $C$ from node $i$ is the product of both probabilities  $k_i^\mathrm{out}(C)/\sum_{i \in C} k_i$ and the probability to leave $C$ from any node (escape probability) is
$P_{\mathrm{esc}}(C) = \sum_{i \in C} k_i^\mathrm{out}(C)/\sum_{i \in C} k_i$.

Escape probability $P_{\mathrm{esc}}(C)=k_\mathrm{out}(C)/k(C)$ equals \textit{conductance} of $C$ for $k(C) < m$ with $m$ the number of all links \cite{fortunato2010community}. For $k(C) > m$ cut $k_\mathrm{out}(C)$ is normalised by $k(V-C)$ (with $V$ the set of all vertices) because subgraphs larger than half the whole graph tend to have smaller cuts $k_\mathrm{out}(C) = k_\mathrm{out}(V-C)$. A smoother normalisation which takes this tendency into account is achieved in \textit{normalised cut} defined by \citeN{shi_normalized_2000} as
\begin{equation}
\Phi(C) =  \frac{k_\mathrm{out}(C)}{k(C)} + \frac{k_\mathrm{out}(C)}{k(V-C)}.
\end{equation}
In the case of link communities we have to cut not links but nodes to separate a link set $L$ from the rest of the graph. \textit{Normalised node-cut} $\Psi(L)$ is a measure of separation of link communities derived from normalised cut $\Phi(C)$ by \shortciteN{havemann_memetic_2017}. It is given by 
\begin{equation}
\Psi(L)= \frac{\sigma(L)}{k_\mathrm{in}(L)} + \frac{\sigma(L)}{k_\mathrm{in}(E-L)}, \quad \mathrm{with} \quad  \sigma(L)    = \sum_{i = 1}^n\frac{k_i^\mathrm{in}(L)  k_i^\mathrm{out}(L)}{k_i},
\label{eq:Psi}
\end{equation} 
where  $i$ runs through all $n$ nodes but $k_i^\mathrm{in}(L)=0$ for all nodes which are not attached to a link in $L$. Set $E$ includes all  $m$ edges and $n$ is the number of all nodes. Note, that $\sigma(L)=\sigma(E-L)$ and $k_\mathrm{in}(E-L)=2m-k_\mathrm{in}(L)$.  \citeN{evans2009line} introduced a random walker who jumps from a link to one of its nodes with probability 1/2 and then chooses one of the links attached to this node. The ratio $\sigma(L)/k_\mathrm{in}(L)$ equals the escape probability of such a \textit{link-node-link random walker}: The probability of a link-node-link random walker to start at any link in  set $L$ and to arrive on node $i$ is $p_i(L)=k_i^\mathrm{in}(L)/\sum_{i = 1}^n k_i^\mathrm{in}(L).$ That means, his probability to leave $L$ from $i$ is 
$p_i(L)k_i^\mathrm{out}(L)/k_i$
and the escape probability is $P_\mathrm{esc}(L) =  \sigma(L)/k_\mathrm{in}(L),$
where $k_\mathrm{in}(L) = \sum_{i=1}^n k_i^\mathrm{in}(L)$ and $\sigma(L)=\sum_{i=1}^n k_i^\mathrm{in}(L)k_i^\mathrm{out}(L)/k_i$ (cf.\,Eq. \ref{eq:Psi}). 
\footnote{This is a new derivation of 
function $\sigma(L)$ used by us 
 for defining 
normalised node-cut $\Psi(L)$ 
which now appears as $\Psi(L) = P_\mathrm{esc}(L) + P_\mathrm{esc}(E-L)$ with $\mathrm{max}(\Psi(L)) = 2$ and not 1, as stated by us. Both probabilities reach a maximum of 1 for a ring graph where each second link is in $L$.}

\bibliographystyle{chicago}
\bibliography{networks}
\end{document}